# Direct Visualization and Manipulation of Tunable Quantum Well State in Semiconducting Nb$_2$SiTe$_4$

*Jing Zhang[1,2,‡], Zhilong Yang[3,‡], Shuai Liu[1,2,4], Wei Xia[1,5], Tongshuai Zhu[3], Cheng Chen[1,6], Chengwei Wang[1,2,7], Meixiao Wang[1,5], Sung-Kwan Mo[6], Lexian Yang[8], Xufeng Kou[5,9], Yanfeng Guo[1,\*], Haijun Zhang[3,\*], Zhongkai Liu[1,5,\*], Yulin Chen[1,5,8,10,\*].*

[1]School of Physical Science and Technology, ShanghaiTech University, Shanghai 201210, China

[2]University of Chinese Academy of Sciences, Beijing 100049, China

[3]National Laboratory of Solid-State Microstructures, School of Physics and Collaborative Innovation Centre of Advanced Microstructures, Nanjing University, Nanjing, 210093, China

[4]Shanghai Institute of Optics and Fine Mechanics, Chinese Academy of Sciences, Shanghai 201800, China

[5]ShanghaiTech Laboratory for Topological Physics, Shanghai 201210, China

[6]Advanced Light Source, Lawrence Berkeley National Laboratory, Berkeley, CA 94720, USA

[7]Shanghai Institute of Microsystem and Information Technology, Chinese Academy of Sciences, Shanghai 200050, China

[8]State Key Laboratory of Low Dimensional Quantum Physics, Department of Physics, Tsinghua University, Beijing 100084, China

[9]School of Information Science and Technology, ShanghaiTech University, Shanghai 201210, China

[10]Department of Physics, University of Oxford, Oxford, OX1 3PU, UK

[‡]These authors contributed equally to this work.

[*]Email address: guoyf@shanghaitech.edu.cn, zhanghj@nju.edu.cn, liuzhk@shanghaitech.edu.cn, yulin.chen@physics.ox.ac.uk

**ABSTRACT:** Quantum well states (QWSs) can form at the surface or interfaces of materials with confinement potential. They have broad applications in electronic and optical devices such as high mobility electron transistor, photodetector and quantum well laser. The properties of the QWSs are usually the key factors for the performance of the devices. However, direct visualization and manipulation of such states are in general challenging. In this work, by using angle-resolved photoemission spectroscopy (ARPES) and scanning tunneling microscopy/spectroscopy (STM/STS), we directly probe the QWSs generated on the vacuum interface of a narrow band gap semiconductor $Nb_2SiTe_4$. Interestingly, the position and splitting of QWSs could be easily manipulated *via* potassium (K) dosage onto the sample surface. Our results suggest $Nb_2SiTe_4$ to be an intriguing semiconductor system to study and engineer the QWSs, which has great potential in device applications.

Quantum well states (QWSs) can form when carriers are tightly confined to a potential well whose thickness becomes comparable to the de Broglie wavelength of the carriers, leading to energy bands with discrete energy levels (*i.e.*, sub-bands). The dimensionally confined electron gas can exhibit exotic quantum phenomena, for instance the high-mobility transport in metal oxide semiconductor field effect transistors (MOSFETs),[1-3] superconductivity and magnetism in LaAlO$_3$/SrTiO$_3$ interface,[4-6] quantum spin Hall effect in HgTe/CdTe[7-9] and InAs/GaSb interfaces,[10, 11] and spin-momentum locking in strong topological insulators.[12-14] On the other side, QWSs are also widely used in electronic industry, including in high electron mobility transistors (HEMTs), infrared detectors, diode laser *etc.*[15-19]

Over the past decades, countless efforts have been invested in generating different types QWSs devices and improving their qualities. The QWSs are generally obtained in thin film structure especially in metallic thin films (such as Bi,[20, 21] Pb,[22, 23] Ag,[24] Cu,[25] Fe,[26] *etc.*), transition metal dichalcogenides/oxides films such as MoS$_2$[27, 28] and SrVO$_3$[29] as well as in thin-film heterostructures (*i.e.*, SrTiO$_3$/LaAlO$_3$[30]). The QWSs could also form on the surface of bulk material (*e.g.*, Bi$_2$Se$_3$,[31, 32] CeCoIn$_5$[33]) which is quite rare. Due to the different geometry, the shape of quantum well and characteristics of the QWSs (spatial extension, energy level separations, evolution into the bulk electronic states, *etc.*) are fairly different in bulk samples comparing to the thin films. Further, the stability and ease of quality control (thickness, doping, *etc.*) on bulk crystals make them appealing for device applications.

On the other hand, the features of the QWSs are the key factors that determine the intriguing physical properties and the performance of the devices.[20, 28-30, 32] For example, the splitting of the sub-bands determines the wavelength of the lasing/absorbing photons, while the carrier concentration, effective mass and mobility strongly affect the performance of HEMTs. The direct visualizing and manipulating these states are also very important to widen our understanding on the fundamental properties of this phenomenon. Previously, the QWSs are usually probed by transport or optical absorption, which are indirect methods and the interpretation heavily relies on the models.[34-37] Recently, the angle-resolved photoemission spectroscopy (ARPES) serves as a powerful tool to directly visualize the electronic structure in the momentum space, thus a direct technique to characterize physical properties of the QWSs (such as effective mass, bandwidth, Fermi velocities, band positions). ARPES has been successfully used in the observation of the strong spin-orbit coupling QWSs formed on the $Bi_2Se_3$/$Bi_2Te_3$ surface,[31] QWSs with large electron-phonon coupling on the surface of $SrTiO_3$ and black phosphorus.[38-40] Meanwhile, scanning tunneling microscopy/spectroscopy (STM/STS) provides the local density of state (DOS) of the QWSs. Together they provide a comprehensive knowledge of the QWSs.

The recently reported semiconductor $Nb_2SiTe_4$ is a semiconductor with a narrow band gap of 0.39 eV and stable in air even in the few-layer limit, making it a candidate for ambipolar devices and mid-infrared response (MIR) detection[41] (typical system with such bandgap size is usually unstable, *e.g.*, black phosphorous). The ambipolar transport behavior with a similar magnitude of electron and hole current and high mobility exhibits high electron mobility up to 1775 $cm^2$ $V^{-1}$ $s^{-1}$ in the monolayer form,[42] as well as the clear response to the MIR wavelength.

In this work, we present the direct visualization of naturally formed QWSs on the top cleavage surface (vacuum-semiconductor interface) of $Nb_2SiTe_4$ and its detailed physical properties extracted (*e.g.*, the energy positions, effective mass and sub-band splitting) by using the ARPES/STM/STS techniques. In addition, we managed to manipulate the QWSs by shifting the QWSs' energy positions and tuning the sub-band splitting, through *in situ* potassium (K) dosage on the surface (an alternative way of QWSs manipulation comparing with the previous gating,[43] substrate interface changing,[44] surface alloy deposition (*e.g.*, Pb[45]) or molecule adsorption methods such as $N_2$,[46] organic molecules and *etc.*[47]). The measured dispersion and evolution of the QWSs show nice agreement with our first-principles calculations. Our study reveals the existence of QWSs in a narrow gap semiconductor, and suggests a pathway to characterize and engineer the QWSs in various systems with important device applications, *e.g.*, photodetectors with tunable photon energies.

**RESULTS AND DISCUSSIONS**

**Formation of QWSs on a semiconductor surface.** The schematics of the QWSs forming on the vacuum semiconductor interface were illustrated in Figure 1. Due to the breaking of translational symmetry, an electric field builds up at the interface alongside with the bending of the conduction and valence bands. The electric fields form a surface quantum well confined to the sample surface, where the QWSs with several quantized sub-bands resides. The dispersion of such type of QWSs could be viewed as the quantized wavefunctions of electrons near $E_F$, as illustrated in Figure 1a(i), b(i). The typical band structure of the QWSs in semiconductors would appear dispersive along the $K_\parallel$ ($K_x$ and $K_y$) directions, while being relatively flat along the $K_\perp$ ($K_z$) direction due to the surface confinement (Figure 1c(i)).

The surface band bending plays a critical role in the formation of the QWSs. Therefore, the QWSs could be modified by changing the surface conditions, *e.g.*, by decorating the surface with K atoms (Figure 1a(ii)). The deposited K provides extra charges onto the sample surface, increasing the electric field and the extent of the band bending, thus steepens the surface potential well, as demonstrated in Figure 1b(ii). The confined QWSs are expected to evolve inside the modified potential well, with the increased splitting of the sub-bands (Figure 1c(ii)).

**Crystal structure and characterizations of $Nb_2SiTe_4$.** The formation and manipulation of this type of QWSs can be realized in $Nb_2SiTe_4$, a narrow gap semiconductor (gap size ~ 0.39 eV) recently discovered.[41] $Nb_2SiTe_4$ crystallizes in a monoclinic lattice with space group $P121/c1$ (No. **14**) with a layered structure shown in Figure 2a(i). Each layer is formed by the chain-like structure (Figure 2a(ii)) along the *a*-axis and the layers are bounded by van der Waals force. The surface topography on the cleaved surface of $Nb_2SiTe_4$ with STM is illustrated in Figure 2b, showing nice arrangement of the topmost tellurium (Te) atoms on the (001) cleaved surface plane. The measured lattice constants obtained from the atomic resolved image are $a = 6.70$ Å, $b = 8.27$ Å, respectively. The high quality of the samples was verified by X-ray diffraction and X-ray photoemission spectroscopy (see the supporting information, Figure S1a, b).

The overall electronic band structure of $Nb_2SiTe_4$ is illustrated in Figure 2d (the definition of the Brillouin zone (BZ) and its projection on the (001) surface are shown in Figure 2c). From the three-dimensional plot of the photoemission spectra, as well as the dispersions along the high symmetry $\bar{Z} - \bar{\Gamma} - \bar{Z}$ (Figure 2e(i)) and $\bar{A} - \bar{\Gamma} - \bar{A}$ directions (Figure 2e(iii)), one can clearly identify an indirectly gapped semiconducting electronic structure. The valence band top is located between the $\bar{A}$ and $\bar{\Gamma}$ points while the conduction band bottom appears between the $\bar{Z}$ and $\bar{\Gamma}$ points, with the indirect gap size measured to be ~ 0.3 eV (Figure S2a). The measured electronic structure shows nice agreement with the first-principles calculations (Figure 2e(ii), (iv)) and the previous STS measurement.[41]

Interesting, several discrete bands are spotted inside the valence bands (as indicated by the arrows in Figure 2e). These discrete bands locate within the bulk band continuum and do not appear in the calculations of the bulk electronic states. Rather, these features are the manifestation of the sub-bands of the QWSs formed on the Nb$_2$SiTe$_4$ cleavage surface.

**QWSs in the valence band.** The QWSs could also be observed in the dI/dV curve measured by the STS (Figure 3b), where the energy intervals match well with the photoemission spectrum near $\bar{\varGamma}$ (the second derivative illustrated in Figure 3a, with the integrated energy distribution in Figure 3b), in spite of a ~200 meV global energy shift, which may come from the inhomogeneity of the sample and different probe range of STM and ARPES. Therefore, the features are excluded from the artifact caused by the photoemission matrix element or the finite $K_z$ resolution (which would bring together the dispersion at other $K_z$ values). In order to verify the QWSs nature of these features, we further carried out photon-energy-dependent ARPES measurement, which allows us to extract the dispersion of band structure along the $K_z$ ($\bar{B} - \bar{\varGamma} - \bar{B}$) direction. From the extracted energy distribution curve (EDC) plot at $K_x = K_y = 0$ Å$^{-1}$ measured at different photon energies (Figure 3d) we find that while the intensity corresponding to the bulk bands shows clear evolution with the changing photon energies (*i.e.*, along the $K_z$ direction), the intensity of the QWSs sub-bands shows little variation, proving their two-dimensional nature (the peaks aligned by the red dotted lines).

To further understand the nature of the QWSs, we calculated the band structure of the QWSs by combining the first-principles calculation with the Poisson's equation of the one-electron surface potential *V(z)*, which is presented in detail in the method section. The calculated QWSs show multiple sub-bands with reducing energy splitting (Figure 3c), in general agrees well with the ARPES result.

**Evolution of the QWSs with K dosing.** As the surface potential $V(z)$ plays a crucial role in the formation of the QWSs, increasing the $V(z)$ by adding additional electrons through the surface dosage of K appears to be a good way to manipulate the QWSs in $Nb_2SiTe_4$. Detailed information about the setup of the K dosage can be found in Figure S3. With increasing K dosage, a gradual downward shift of valence and conduction bands (Figure 4a-c along the high symmetry directions $\bar{A} - \bar{\Gamma} - \bar{A}, \bar{Z} - \bar{\Gamma} - \bar{Z}$, and the second derivative of the later, respectively) was clearly observed, suggesting the increase of the extent of the band bending. The sub-bands of the QWSs appear to depart from each other while shifting towards higher binding energies together (additional information can be found in Figure S2b). Such systematic evolution with the surface K dosage (increasing of the depth of the surface potential well ($V_{ss}$)), could be repeated by our theoretical calculation (Figure 4d), which shows the same downward shift of the sub-bands and the increase of the energy intervals.

The evolution of QWSs from both the experiment and the theoretical calculation are summarized statistically in Figure 5. We plot the energy positions of the first three sub-bands as a function of the calculated K dosage level, and a similar trend can be found between the experiment (black axes) and calculation (blue axes). If we rescale the energy axis and shift the relative energy position, the two results can be well aligned. The difference here is likely due to the fact that parameters used in our calculation cannot thoroughly capture the real conditions of the material surface. Nevertheless, our simple model already quantitatively reproduced the formation and manipulation of the QWSs.

Our observation of the QWSs in $Nb_2SiTe_4$ provide a revenue for the investigation on this material system. As the QWSs appear in the form of flat bands along all directions near $\bar{\Gamma}$, they create large density of state at certain energy levels (0.3 to 1.2 eV from the bottom of the conduction bands), which may lead to strong photon absorption/emission. In fact, $Nb_2SiTe_4$ already demonstrates strong MIR.[41] More importantly, the tunable QWSs provide the possibility to select the wavelength of the absorbed/emitted photons, therefore have great application potentials in the development of future optoelectronic devices.

We note that in addition to the K dosage in the vacuum environment, the surface potential wells which are critical to the formation of the QWSs could be created in alternative ways, such as fabricating heterostructures with materials possessing different chemical potential or applying gate voltage to the thin film of $Nb_2SiTe_4$. These methods could apply to the normal environment and compatible with device fabrications. For sure, different approaches for tuning the electronic structure create different physical environments for the QWSs, but the experiment carried in the UHV environment would illustrate the key physics that would hint for tuning the electronic structure in ambient environment, device design and applications.

**CONCLUSIONS**

In conclusion, we systematically investigated the electronic structure of the recently discovered interesting semiconductor, $Nb_2SiTe_4$, and confirmed the indirect band gap (~ 0.3 eV). Naturally formed QWSs were observed on the cleavage surface of $Nb_2SiTe_4$, with discrete sub-bands inside the valence band. Through *in situ* surface K dosage, we are able to tune the surface potential and therefore change the positions and the splitting of sub-bands of the QWSs. Comparing to the previous reports on QWSs (*e.g.*, in $Bi_2Se_3$[31, 32]), the QWSs in $Nb_2SiTe_4$ show the following fascinating properties. Firstly, the sub-bands of QWSs appear as flat-bands along all directions near $\bar{\Gamma}$, showing contrast to the other QWSs observed (*e.g.*, $Bi_2Se_3$,[31, 32] Bi,[20, 21] Pb,[22, 23] *etc*.) which exhibit fairly dispersive bands. The flat sub-bands create peaks in the DOS of $Nb_2SiTe_4$, and allow absorption/emitting of photons with selective photon energy. Besides, the QWSs are observed in the valence band, suggesting a *p*-type QWS, which is rarely studied and reported (the only case is in the $Bi_2Se_3$,[31, 32] $CeCoIn_5$[33]) due to their low mobility and difficulty to identify them from the transport measurement. Our results not only reveal the fascinating QWSs on the surface of a narrow band gap semiconductor, which has great application potential, but also proposes an intriguing method to probe and manipulate the QWSs, which is an alternative way to widen our fundamental understanding of this interesting phenomenon.

**METHODS**

**Crystal growth.** Mixing powder in a mortar for 20 mins with a molar ratio of 1:2:8 of the Nb (niobium, 99.95%), Si (silicon, 99.999%) and Te (tellurium, 99.99%), the completely mixed powder was placed in alumina crucible and sealed using a quartz tube at ~$10^{-4}$ Pa. Then the tubes are healed up to 1150 ˚C in 15 h and stay for 5 h, then ramp down to 750 ˚C with a rate of 1 ˚C h$^{-1}$ and stay for another 15 h in a high temperature well-type furnace. Extra power Si and Te will play a role of flux in reducing the melting temperature of Nb and they will be quickly removed using a centrifuge with vacuum of ~$10^{-4}$ Pa keeping the temperature stage at 750 ˚C. The final step is to cool down the quartz tube to room temperature in the air and then $Nb_2SiTe_4$ are obtained with metallic luster flakes.

**STM and STS.** STM used in this work is a commercial product provided by Unisoku cooperation. $Nb_2SiTe_4$ single crystal was cleaved in the preparation chamber and transferred *in situ* to the STM sample stage connected to a cryogenic stage kept at 77.8 K (by liquid nitrogen) for measurement under ultra-high vacuum (UHV) below $2 \times 10^{-10}$ Torr. Silver islands grown on *p*-type Si (111)-7 × 7 were used to calibrate the Pt-Ir tips, which were used for both imaging and tunneling spectroscopy measurements. And the dI/dV curves are amplified by Lock-in with reference signal frequency of 997.233 HZ and amplitude of 5 mV.

**ARPES.** Synchrotron based ARPES measurements were performed at beam line 10.0.1 of the Advanced Light Source (ALS). The samples were cleaved *in situ* and measurements were performed under UHV below $3 \times 10^{-11}$ Torr at ALS. Data were collected by a Scienta R4000 analyzer. The total energy and angle resolutions were 10 meV and 0.2°, respectively.

Home-built helium-lamp-based ($h\nu = 21.2\ eV$) ARPES measurements were performed at ShanghaiTech University. The samples were cleaved *in situ* and measured under UHV below $4 \times 10^{-11}$ Mbar. Data were collected by a DA30L analyzer. The total energy and angle resolutions were 20 meV and 0.2°, respectively. All ARPES measurements were performed at liquid nitrogen temperature.

**First-principles band structure calculations.** The first-principles calculations were carried out in the framework of the Perdew-Burke-Ernzerhof-type generalized gradient approximation of the density functional theory through employing the VASP package with the plane-wave pseudo-potential method.[48-50] The kinetic energy cutoff is fixed to 450 eV and the *k*-point mesh is taken as $\Gamma$-centered $8 \times 8 \times 8$ grid for the bulk calculations. The experimental crystal structure is taken and atomic positions are fully relaxed until the atomic force on each atom is less than $10^{-2}$ eV Å$^{-1}$. The spin-orbit coupling is included in the whole calculations. Tight-binding Hamiltonian based on Wannnier functions is constructed by projecting the Bloch states onto the Nd-4d, Te-5p orbitals.[51, 52] The surface states are calculated by the iterative Green's function method using WannierTools package.[53]

The one-electron potential *V(z)*, describing the band bending in the space-charge region as a function of depth $z$ below a semiconductor surface, must satisfy Poisson's equation: $\frac{d^2V}{dz^2} = -\frac{e}{\varepsilon_r \varepsilon_0}(-N_D + N_A - p(z) + n(z))$ with the boundary conditions: $V(z) \to 0$ as $z \to \infty$, $\frac{dV}{dz}|_{z=0} = \frac{e}{\varepsilon(0)\varepsilon_0} N_{ss}$, where, $N_D$, $N_A$ are the bulk donor/acceptor density, assumed constant throughout the semiconductor, and $p(z), n(z)$ are the electron/hole density and $N_{ss}$ is the surface state density. $n(z) = \int_0^\infty g_c(E) f_{FD} dE$, $p(z) = \int_{-\infty}^{E_v} g_v(E)(1 - f_{FD}) dE$, $f_{FD}(E) = \frac{1}{1+e^{\beta[E-E_F+V(z)]}}$ where $g_c(E), g_v(E)$ are the density of states for the conduction and valence bands, respectively, and we get them from first-principles calculations. Here we set $V_{ss}$ instead of $N_{ss}$ as initial boundary conditions.

A 50-unit cell-thickness tight-binding supercell Hamiltonian, based on maximally localized Wannier functions, was constructed from first-principles calculations. In order to consider the electric-field effect, the resulting potential *V(z)* was then added to the onsite energy of the supercell tight-binding Hamiltonian.

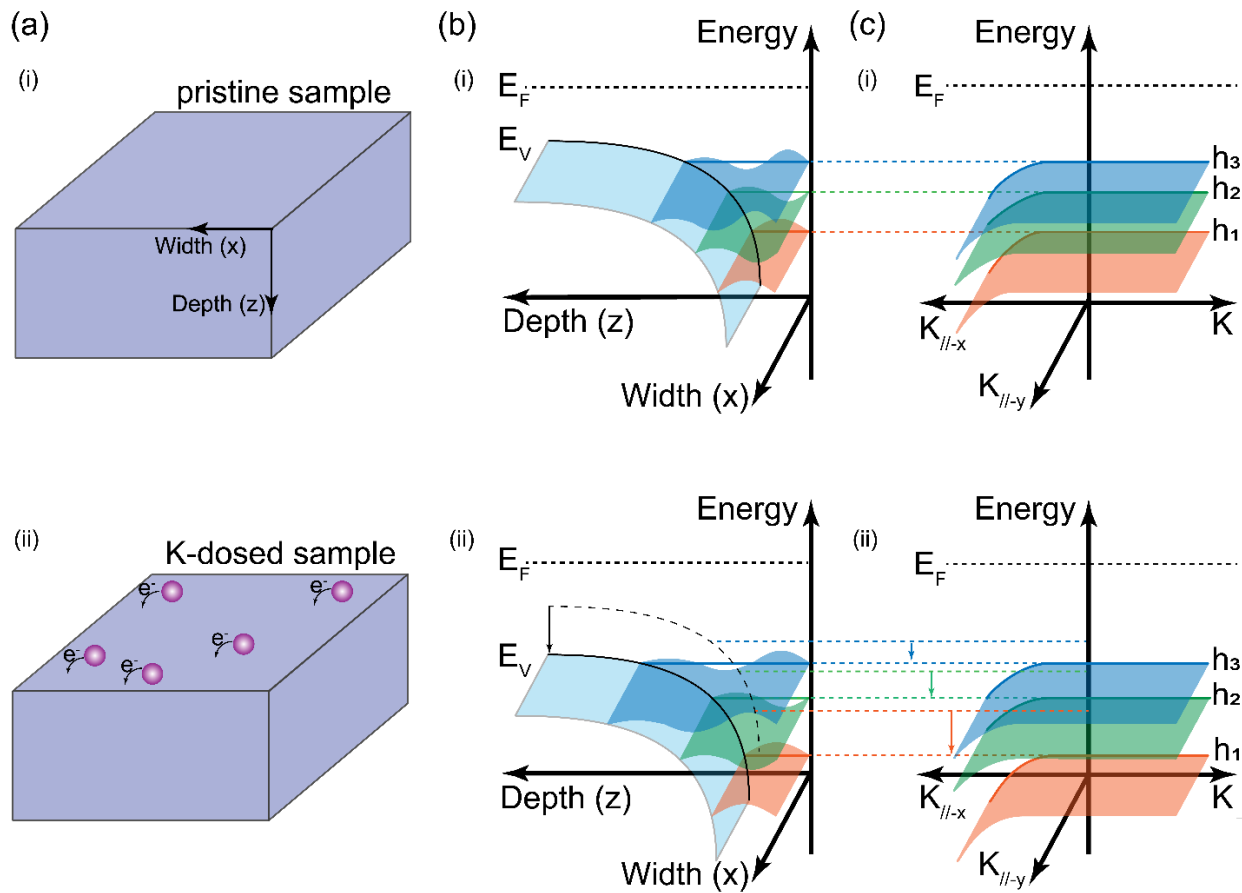

**Figure 1. The formation of QWSs on a semiconductor surface. (a)** Schematic of the sample surface before and after K-dosage. **(b)** Schematic of QWSs at the sample surface before and after K-dosage. **(c)** Schematic of QWSs along parallel momentum $K_{//}$ and vertical momentum $K_{\perp}$, $E_F$: Fermi energy. $E_v$: surface potential energy.

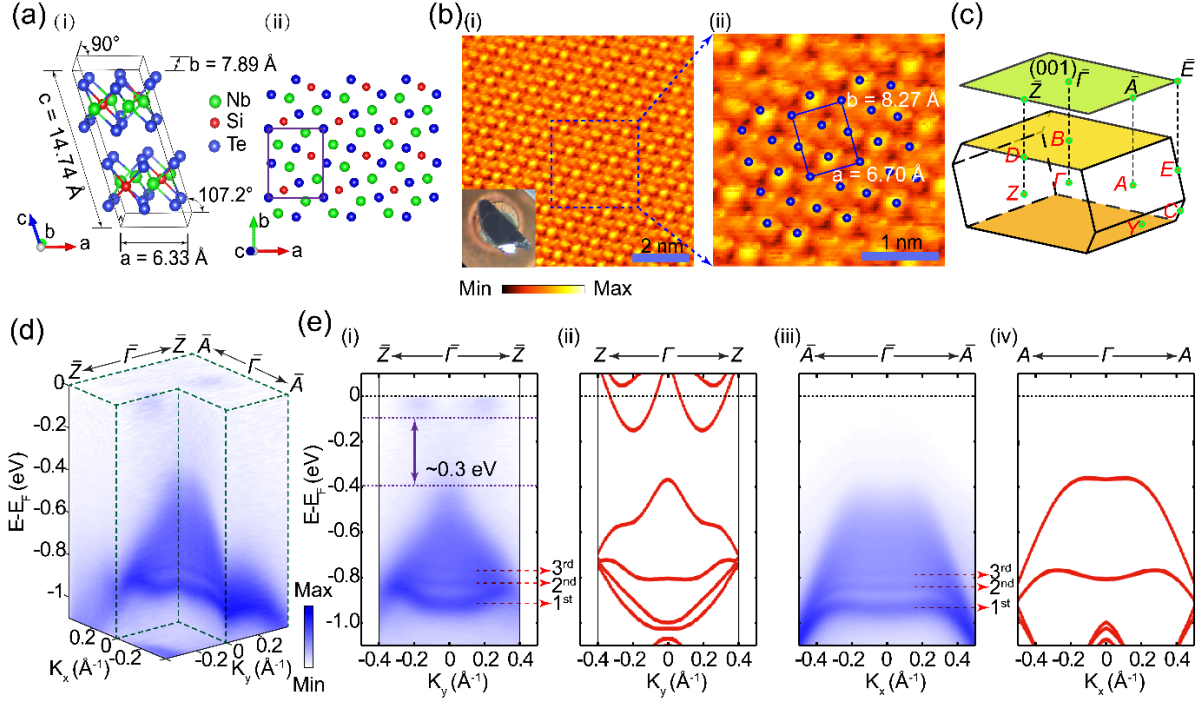

**Figure 2. Crystal structure and basic characterizations of Nb$_2$SiTe$_4$. (a)** Schematic of (i) the layered monoclinic crystal structure of Nb$_2$SiTe$_4$, (ii) lattice structure in the *ab* plane with one surface unit cell labelled by the blue lines. **(b)** (i) Surface topography of the (001) surface of Nb$_2$SiTe$_4$ on the 8 nm × 8 nm sample surface, with sample bias $U_s$ = 40 mV and tunneling current $I_s$ = 500 pA. The inset shows the picture of a high quality Nb$_2$SiTe$_4$ single crystal. (ii) Zoomed-in topography scan of an area in (i). Identified Te atom of (001) crystal plane is marked with one surface unit cell labelled by the blue lines. **(c)** Schematic of bulk Brillouin zone and its projection to the (001) surface with high symmetry points labelled. **(d)** 3D Volume plot of the overall band structure acquired by helium-lamp-based ($hv = 21.2\ eV$) ARPES. **(e)** Photoemission spectra along the $\bar{Z} - \bar{\Gamma} - \bar{Z}$ (i) and $\bar{A} - \bar{\Gamma} - \bar{A}$ (iii) directions, respectively. Red arrows indicate the identified sub-bands of QWSs. An indirect band gap of ~ 0.3 eV is labelled. The results are compared with the theoretical calculations of the bulk band structure along the $Z - \Gamma - Z$ (ii) and $A - \Gamma - A$ directions (iv), respectively. Data were collected at $hv = 21.2\ eV$.

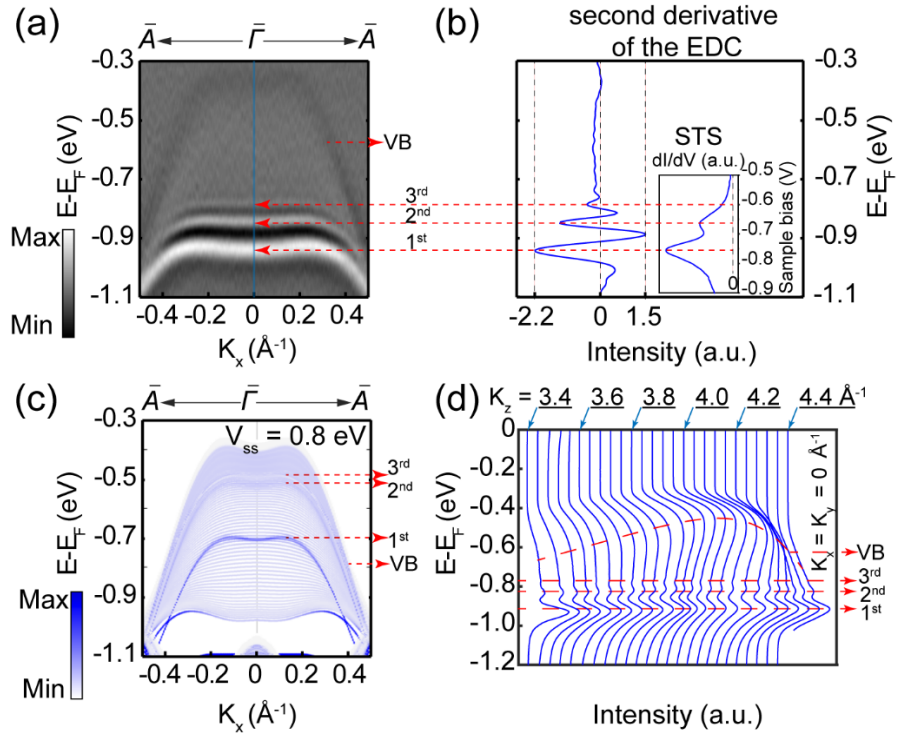

**Figure 3. QWSs in the valence band. (a)** Plot of the second derivative of the photoemission intensity along the $\bar{A} - \bar{\Gamma} - \bar{A}$ direction. Sub-bands of the QWSs and valence bands are labelled. Data were acquired at $h\nu = 21.2\ eV$. **(b)** Extracted second derivative of EDC from (a) near $\bar{\Gamma}$ (indicated by the blue line in (a)). The inset shows dI/dV spectrum on $Nb_2SiTe_4$ measured by STS at liquid nitrogen temperature with the same energy range. Red arrows indicate the identified sub-bands of QWSs. **(c)** Calculated band dispersions along the $\bar{A} - \bar{\Gamma} - \bar{A}$ direction with $V_{ss}$ = 0.8 eV. Arrows indicate the identified QWSs and bulk valence band. **(d)** Stacked plot of EDCs at $K_x = K_y = 0$ Å$^{-1}$ but different $K_z$ values extracted from photon energy dependent measurement whose photon energy ranges from 37 to 69 eV. Arrows and dashed lines mark the positions of the valence band and QWSs.

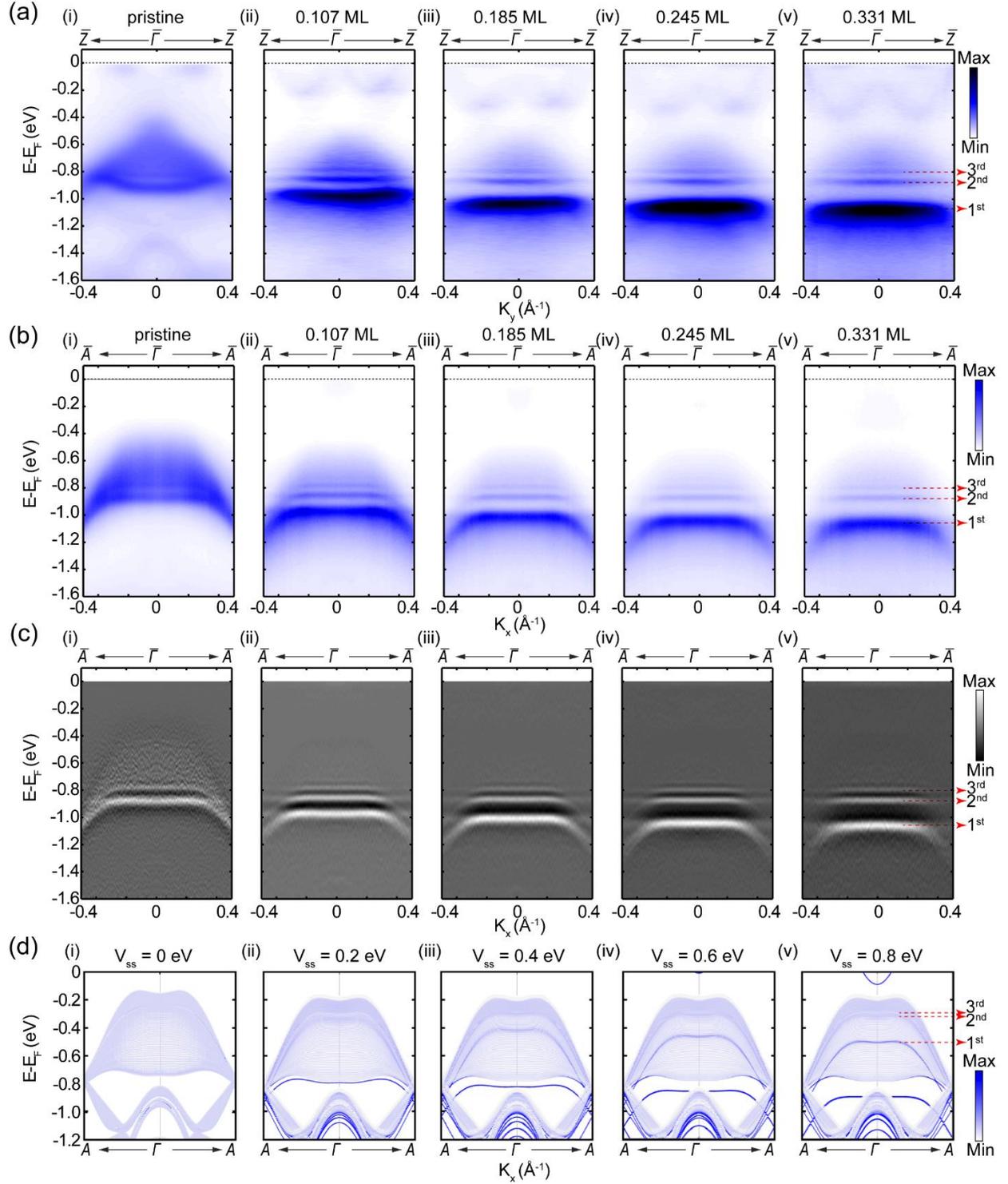

**Figure 4. Evolution of the QWSs with K dosing. (a) (i-v)** Evolution of band dispersion along the $\bar{Z} - \bar{\Gamma} - \bar{Z}$ direction at different K dosage. **(b) (i-v)** Evolution of band dispersion along the $\bar{A} - \bar{\Gamma} - \bar{A}$

directions at different K dosage. **(c)** Second derivative analysis results of ARPES intensity in (b). **(d)** Calculated QWSs band structure at different $V_{ss}$ values. The first three sub-bands are labeled by red arrows in (a-d). Data of (a) and (b) were acquired at $h\nu = 21.2\ eV$. The estimation of the coverage of K dosage in units of monolayer (ML) are discussed in details in the supporting information. We could also note the reduction of the bandgap in (a). The physical origin and its application are discussed in the supporting information.

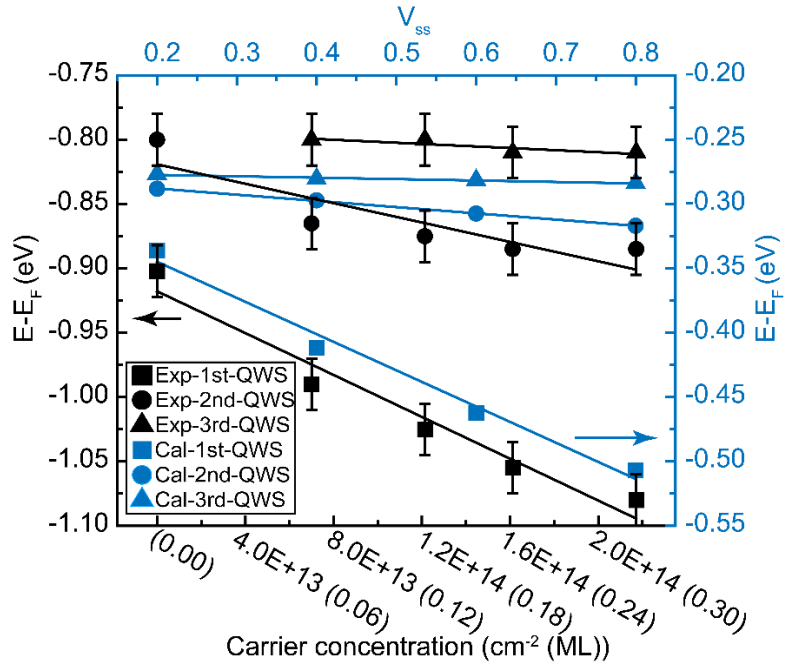

**Figure 5. Summary of QWSs evolution with surface potential.** The black markers are the extracted energy positions of the first three sub-bands of the QWSs from ARPES measurement whereas the blue marks are the extracted energy positions of the first three sub-bands of the QWSs from the calculation results. The estimation of the carrier concentration from K dosage is explained in detail in the supporting information. The energies are scaled for a clear comparison of the trend of evolution.

## ASSOCIATED CONTENT

**Supporting Information**.

The Supporting Information is available free of charge at

Additional figures and data to support the results in the main text: Powder X-ray diffraction and X-ray photoelectron spectroscopy; Indirect band gap of the pristine sample and evolution of QWSs with K dosing; Setup of vacuum K dosing and sample cleavage; Evolution of the constant energy contour of $Nb_2SiTe_4$ at $E_F$ with K dosing and estimation of the monolayer; Evolution of XPS spectra of K dosed $Nb_2SiTe_4$; Projected QWSs structure at different unit layer. Band structure evolution with surface potential from two path *A-G-A* and *Z-G-Z*. (PDF)

**AUTHOR INFORMATION**

**Corresponding Author**

**Yanfeng Guo** – School of Physical Science and Technology, ShanghaiTech University, Shanghai 201210, China; ShanghaiTech Laboratory for Topological Physics, Shanghai 201210, China

Email: guoyf@shanghaitech.edu.cn

**Haijun Zhang** – National Laboratory of Solid-State Microstructures, School of Physics and Collaborative Innovation Centre of Advanced Microstructures, Nanjing University, Nanjing, 210093, China

Email: zhanghj@nju.edu.cn

**Zhongkai Liu** – School of Physical Science and Technology, ShanghaiTech University, Shanghai 201210, China; ShanghaiTech Laboratory for Topological Physics, Shanghai 201210, China

Email: liuzhk@shanghaitech.edu.cn

**Yulin Chen** – School of Physical Science and Technology, ShanghaiTech University, Shanghai 201210, China; ShanghaiTech Laboratory for Topological Physics, Shanghai 201210, China; State Key Laboratory of Low Dimensional Quantum Physics, Department of Physics, Tsinghua University, Beijing 100084, China; Department of Physics, University of Oxford, Oxford, OX1 3PU, UK

Email: yulin.chen@physics.ox.ac.uk

**Author**

**Jing Zhang** – School of Physical Science and Technology, ShanghaiTech University, Shanghai 201210, China; University of Chinese Academy of Sciences, Beijing 100049, China

**Zhilong Yang** – National Laboratory of Solid-State Microstructures, School of Physics and Collaborative Innovation Centre of Advanced Microstructures, Nanjing University, Nanjing, 210093, China

**Shuai Liu** – School of Physical Science and Technology, ShanghaiTech University, Shanghai 201210, China; University of Chinese Academy of Sciences, Beijing 100049, China; Shanghai Institute of Optics and Fine Mechanics, Chinese Academy of Sciences, Shanghai 201800, China

**Wei Xia** – School of Physical Science and Technology, ShanghaiTech University, Shanghai 201210, China; ShanghaiTech Laboratory for Topological Physics, Shanghai 201210, China

**Tongshuai Zhu** – National Laboratory of Solid-State Microstructures, School of Physics and Collaborative Innovation Centre of Advanced Microstructures, Nanjing University, Nanjing, 210093, China

**Cheng Chen** – School of Physical Science and Technology, ShanghaiTech University, Shanghai 201210, China; Advanced Light Source, Lawrence Berkeley National Laboratory, Berkeley, CA 94720, USA

**Chengwei Wang** – School of Physical Science and Technology, ShanghaiTech University, Shanghai 201210, China; University of Chinese Academy of Sciences, Beijing 100049, China; Shanghai Institute of Microsystem and Information Technology, Chinese Academy of Sciences, Shanghai 200050, China

**Meixiao Wang** – School of Physical Science and Technology, ShanghaiTech University, Shanghai 201210, China; ShanghaiTech Laboratory for Topological Physics, Shanghai 201210, China

**Sung-Kwan Mo** – Advanced Light Source, Lawrence Berkeley National Laboratory, Berkeley, CA 94720, USA

**Leixian Yang** – State Key Laboratory of Low Dimensional Quantum Physics, Department of Physics, Tsinghua University, Beijing 100084, China

**Xufeng Kou** – ShanghaiTech Laboratory for Topological Physics, Shanghai 201210, China; School of Information Science and Technology, ShanghaiTech University, Shanghai 201210, China

**Author Contributions**

Z.L. and Y.C. conceived the project; J.Z., C.C. and S.K.M. performed the ARPES and XPS study with the help from L.Y.; Z.Y., T.Z. and H.Z. performed the theoretical calculation; S.L. and M.W. performed the STM study; W.X. and Y.G. synthesized the crystals and performed the XRD measurement. All authors contributed to the preparation of the manuscript.

J.Z. and Z.Y. contributed equally to this work.

**Notes**

The authors declare no competing financial interest.

**ACKNOWLEDGMENT**

The authors thank H.Y., H.Z. and A.L. for helpful discussions. We acknowledge the Beamline 10.0.1 of the Advanced Light Source (ALS), beam line BL03U of Shanghai Synchrotron Radiation Facility (SSRF), China, and beam line BL13U of National Synchrotron Radiation

Laboratory (NSRL), China, for accessing and preliminary ARPES studies. The work is supported by the National Key R&D program of China (Grants No.2017YFA0305400). We thank the support from Analytical Instrumentation Center (contract no. SPST-AIC10112914), SPST, ShanghaiTech University. We acknowledge Natural Science Foundation of Jiangsu Province (No. BK20200007), the Natural Science Foundation of China (Grants No. 12074181 and No. 11834006) and the Fok Ying-Tong Education Foundation of China (Grant No. 161006).